# Deep-ultraviolet Layered Oxide $B_2S_2O_9$ with Strong and Robust Second Harmonic Generation

Lei Kang,[1] Xiaomeng Liu,[1] Zheshuai Lin,[1]* Bing Huang[2,3]†

[1] *Technical Institute of Physics and Chemistry, Chinese Academy of Sciences, Beijing 100190, China*
[2] *Beijing Computational Science Research Center, Beijing 100193, China*
[3] *Department of Physics, Beijing Normal University, Beijing 100875, China*
E-mails: zslin@mail.ipc.ac.cn (Z.L.); bing.huang@csrc.ac.cn (B.H.)

**Two-dimensional (2D) layered semiconductors with both ultrawide bandgap and strong second harmonic generation (SHG) are essential for expanding the nonlinear optical (NLO) applications to deep-ultraviolet (DUV) region in nanoscale. Unfortunately, these materials are rare in nature and have not been discovered until now. In this Letter, we predict the $B_2S_2O_9$ (BSO), an existing layered oxide, can exhibit both DUV bandgap and strong SHG effects, comparable to the best known DUV NLO bulks. The strong SHG intensities in BSO, originated from the ordered arrangement of polar $SO_4$ and $BO_4$ tetrahedra forming planar structure, are linearly tunable by the layer thickness. Surprisingly, the spontaneous rotations of rigid tetrahedra under strains can induce the (nearly) zero Poisson's ratios in BSO, which simultaneously result in the robust SHG effects against large strains, fundamentally differing from other known 2D NLO semiconductors. The discovery of BSO may provide an unprecedented opportunity to explore DUV NLO physics and applications in 2D limit.**

Two-dimensional (2D) van der Waals (vdW) layered materials may offer unprecedented opportunities to expand or enhance functionalities from 3D bulk down to monolayer (ML) thickness [1-5]. Generally, the energy bandgap ($E_g$) of a material determines its transparency in the short-wave spectrum. From the perspective of device applications, it is highly desired to have complete 2D systems that can cover the entire energy ranges from infrared (IR) to deep-ultraviolet (DUV, $\lambda$<200 nm). In particular, the existence of strong second harmonic generation (SHG) is important for 2D semiconductors in constructing novel nonlinear optical (NLO) devices [5-13]. Until now, several existing systems, *e.g.*, graphene [1], black phosphorus (BP) [3], transition-metal dichalcogenides (TMDs) [4], and hexagonal boron nitride (*h*-BN) [14,15], have been successfully discovered to cover transparent regions from IR to UV. And some of them, *e.g.*, TMDs [5-11] and *h*-BN [5,11,12], can exhibit strong SHG intensities, which are, however, quite sensitive to the external perturbations [11-13, 16-19]. Unfortunately, none of existing 2D materials proposed so far has a sufficiently large DUV $E_g$, which in turn limits the DUV NLO applications in 2D limit [20].

The challenges to discover 2D DUV semiconductors with strong NLO capacity are twofold. Firstly, it is impractical to directly search for vdW layered DUV materials from material database (>200,000 candidates), as the current understanding on their critical structural characteristics is limited. Secondly, although exploring novel DUV NLO materials has last for >30 years since the discovery of $KBe_2BO_3F_2$ (KBBF) [21], the only practical DUV NLO crystal nowadays, the available candidates are still extremely rare [22,23]. And none of them has a vdW layered structure that could be exfoliated into ML. In addition, the phase-matching (PM) condition, critical for NLO conversion, is too stringent to be satisfied for most existing DUV crystals [23,24]. Fortunately, the 2D DUV NLO semiconductors would naturally eliminate the PM requirement as their thickness are far below the coherent wavelength. Therefore, it is highly urgent to discover novel 2D DUV NLO materials in terms of both scientific interests and technological implications.

In this Letter, we have developed an important understanding of the critical structural features of 2D DUV NLO materials. Using the first-principles calculations, we have successfully discovered an existing vdW layered oxide $B_2S_2O_9$ (BSO), formed by the planar ordered arrangement of polar B- and S-centered tetrahedra, can exhibit both ultrawide DUV $E_g$ (>8 eV) and strong SHG intensities (~2.5×KBBF). Interestingly, differing from TMDs and *h*-BN, the SHG intensities of BSO can be linearly tuned by its thickness, due to the intrinsic inversion asymmetry and weak interlayer coupling between layers. Remarkably, under the external strains, the inherent rigid tetrahedra in ML BSO can spontaneously rotate to avoid the tetrahedral distortions, giving rise to the (nearly) zero in-plane Poisson's ratios and minimizing the changes of macroscopic polarization. Consequently, the robust SHG intensities against strains can be achieved in ML BSO, fundamentally differing from other known 2D NLO materials.

All the first-principles calculations are performed by the norm-conserving pseudopotential method based on density functional theory (DFT), as implement in CASTEP [25]. The energy cutoff of 550 eV and Monkhorst-Pack *k*-point meshes of 0.04 Å$^{-3}$ in the Brillouin zone (BZ) are adopted, respectively. All the structures are fully relaxed until the residual forces on each atom is less than 0.01 eV Å$^{-1}$. The semi-empirical vdW correction for DFT is adopted to deal with the weak interlayer interactions in layered systems [26]. All the electronic structures are calculated by hybrid PBE0 functional [27] and the optical properties are calculated by standard scissors-corrected DFT methods (see Supplemental Material [28] for details), which have been established as an efficient *ab initio* way to accurately calculate the $E_g$, linear (*e.g.*, refractive indices $n$) and NLO effects (*e.g.*, SHG coefficients $d_{ij}$) of DUV materials [23,29], especially for borates [22]. The phonon spectra and elastic tensors of BSO systems are calculated using first-principles linear response approach [30] and finite strain technique [31], respectively.

To search for 2D DUV NLO materials, we have proposed three basic procedures, as illustrated in **Fig.1a**, to distill potential candidates: (i) to realize DUV $E_g$: the candidates should exhibit significant cation-anion electronegativity differences; we focus on oxides and halides with sufficiently large energy differences (*e.g.*, >6 eV) between cation and anion atomic orbitals that form the band-edge states. (ii) to realize strong SHG: their structural motifs should include polar anionic units [22,23], *e.g.*, tetrahedra, aligned in a manner to exhibit significant SH polarization ($P$); for example, the ordered $(BO_4)^{5-}$ tetrahedra play an important role in achieving both ultrawide $E_g$ (~9.2 eV) and strong SHG effects (~0.7 pm/V) in $SrB_4O_7$ and $BPO_4$ [22]. (iii) to realize vdW layered structure: the polar tetrahedra could form a planar or quasi-planar framework by sharing their corner anion atoms; meanwhile, there should be no unpaired electrons on the outmost surfaces, avoiding the formation of covalent bonds between neighboring layers. Based on extensive structural search in the inorganic crystal structure database (ICSD), we have successfully discovered an existing oxide, $B_2S_2O_9$ (BSO) [32], could meet all the (i)-(iii) criteria.

As shown in **Fig. 1b**, the AA stacking of BSO ML forms the ground-state structure of bulk BSO ($C_2$ symmetry). The calculated lattice constants (Table S1 [28]) are in good agreement with the experimental data [32]. The large cation B and anion O 2$p$-orbital energy difference (~6 eV) indicates the possibility of achieving DUV $E_g$ in BSO (criteria i). There are three nonequivalent O sites in the BSO system. Interestingly, for each BSO ML, it includes two layers of polar $BO_4$ and $SO_4$ tetrahedra that align in a parallel ordering via sharing corner O (O1) atoms in the *a-b* plane, which could maximize the intrinsic $P$ (approximately along *a-b* diagonal direction, see Fig. S1 [28]) and simultaneously achieve strong SHG effects (criteria ii). Between the two neighboring $(BO_4)^{5-}$-$(SO_4)^{2-}$ layers, the two $BO_4$ units can form one B-O-B bond via sharing the bridge O (O2) atom along the *c* axis. Moreover, Mulliken population analysis (Table S2 [28]) confirm that these outmost S-O surface bonds in ML BSO exhibit more bond population (~0.86) than the others (~0.50), indicating the formation of S=O double-bonds. As a result, there are no unpaired electrons on the surface O (O3) atoms, so that BSO can form a stable vdW layered structure (criteria iii).

In order to understand the feasibility of exfoliating BSO ML, its cleavage energy ($E_{cl}$) is calculated, comparing with those of $MoS_2$ and *h*-BN. As shown in **Fig. 2a**, the calculated $E_{cl}$ ~20 meV/atom in BSO, significantly smaller than those of $MoS_2$ (~40 meV/atom) and *h*-BN (~70 meV/atom) [33], indicating that BSO ML may be easily peel off by the mechanical methods. As shown in **Fig. 2b**, the phonon spectrum of ML BSO is calculated to verify its dynamical stability. Overall, there are no imaginary phonon modes, demonstrating that the ML BSO is dynamically stable, similar to its bulk structure (Fig. S2 [28]). Furthermore, first-principles molecular dynamical (MD) simulations are performed to confirm its thermal stability at room temperature (RT). As shown in **Fig. 2c**, the *a* and *b* lattice constants oscillate in small ranges under thermal fluctuations, indicating that ML BSO can be stable (at least) at RT.

The calculated in-plane Young modulus of the BSO system along *a* and *b* axis are 154 and 104 GPa (Table S3 [28]), respectively, similar to that of MoS$_2$ [4] and sufficiently large for good mechanical robustness. Surprisingly, we find that the in-plane Poisson's ratios of BSO are almost zero ($v_{ab}$=-0.008 and $v_{ba}$=-0.006 [28], denoted as anepirretic [34]), originated from the existence of rigid BO$_4$/SO$_4$ tetrahedra. The high rigidity is mostly contributed by the strong $sp^3$ orbital hybridizations in these tetrahedra [35]. Therefore, in response to a linear-elastic strain along *a* (*b*) axis, the tetrahedra prefer to rigidly rotate to eliminate the tetrahedral distortions (with negligible changes of volumes and bond lengths), maintaining the *b* (*a*) lattice nearly constant. As discussed in the following, this unusual elastic behavior in BSO, the first known 2D anepirretic material, can play a key role in realizing robust SHG effects beyond other 2D NLO materials.

As shown in **Fig. 3a**, ML BSO is a direct-gap semiconductor with a DUV $E_g$~8.6 eV, comparable to that of KBBF (the calculated KBBF $E_g$~8.3 eV, same as the experimental $E_g$~8.3 eV [21,22]). As shown in inset of **Fig. 3b**, the large energy differences (~6 eV) between B 2*p* (S 3*p*) and O 2*p* orbitals induced significant charge transfers, along with the BO$_4$/SO$_4$ tetrahedral-coordination-induced strong $sp^3$ orbital hybridizations, can eventually result in extremely large anion-cation bonding-antibonding separations, forming the band-edge states of ML BSO. This is consistent with the calculated partial density of states (PDOS) in **Fig. 3b**, in which the valence band maximum (VBM) and conduction band minimum (CBM) are mostly contributed by the anion (O) and cation (B/S) *p* orbitals, respectively. Interestingly, when the number (*N*) of BSO layers increases (**Fig. 3d**), the $E_g$ of BSO system slightly decreases from 8.63 to 8.46 (*N*=2) to 8.41 (*N*=3) and to 8.34 (*N*=∞) (Fig. S3 [28]). Meanwhile, the critical thickness ($N_c$) for a direct-indirect $E_g$ transition in BSO systems is close to *N*=∞, differing from those in TMDs, *e.g.*, $N_C$=2 for MoS$_2$ [4,36]. This is because a new VBM away from *Γ*-point emerges only if the band dispersion along *c* direction can be included, contributing to an indirect (quasi-direct) $E_g$.

Due to the parallel orderings of tetrahedra and the existence of intrinsic *P* in BSO, it is expected that its SHG intensity could be strong. Indeed, the calculated three major independent non-vanishing SHG coefficients, *i.e.*, $d_{16}$, $d_{14}$, $d_{23}$, of ML BSO (Table S4 [28]) are 2.2, 0.6 and 2.8 times of KBBF (calculated KBBF $d_{16}$~0.41 pm/V [28], close to the experimental $d_{16}$~0.45 pm/V [21]), respectively. Overall, the average SHG intensity of BSO is ~2.5×KBBF, among the highest values in all known DUV crystals [23,37]. Since the optical transitions between the band-edge states are allowed (Fig. S4 [28]), they may largely contribute to the SHG intensity of ML BSO [22]. As shown in **Fig. 3c**, the band-resolved SHG contributions confirm that the band-edge states make the dominant contributions to the high $d_{23}$ value (see Fig. S5 [28] for a similar result of $d_{16}$). In addition, the SHG-weighted density is employed to intuitively identify the orbitals contributed to $d_{23}$. Interestingly, as shown in **Fig. 3e**, a large amount of SHG density is contributed by the orbitals in the S=O3 and B-O2-B bonds, consistent with their significant PDOS contributions around band edges (**Fig. 3b**). In addition, it notes that although our calculations show that the PM condition for DUV SHG is not satisfied for BSO bulk due to the small birefringence (~0.03, see Table S4 [28]) induced by the less anisotropic tetrahedra

(**Fig. S6** [28]), ML and thin-film BSO without PM requirement are still promising for DUV NLO conversion.

We further investigate the $N$- and strain-dependent SHG effects in BSO, compared to that of $h$-BN and MoS$_2$. Despite the inversion asymmetry ($D_{3h}$) of ML $h$-BN and MoS$_2$, the layer stacking restores inversion symmetry ($D_{3d}$) for even $N$, leading to an oscillatory SHG response. Taking MoS$_2$ as an example (**Fig. 4a**), the calculated $d_{11}$ values oscillate as $N$ increases, agreeing with the experimental observations [11,16,18], which cannot be enhanced by increasing $N$. Oppositely, the existence of inversion asymmetry is independent of $N$ in BSO systems. As shown in **Fig. 4a**, due to the weak interlayer interactions (**Fig. 2a**), it is found that the contribution of each BSO layer to the total SHG intensity is nearly equal, giving rise to a linear $N$-dependence of $d_{23}$ ($d_{16}$). Interestingly, it is estimated that the SHG intensity of trilayer BSO could already be higher than that of ML $h$-BN (Table S4 [28]), given that the $E_g$ of BSO is much larger (> 2 eV) than that of $h$-BN [37].

The calculated strain-dependent $d_{23}$ and $d_{16}$ values are shown in **Fig. 4b**. Generally, a negative (positive) biaxial-strain ($\varepsilon$) can homogenously decrease (increase) the dipole intensity of a 2D NLO system, resulting in a decreased (increased) $d_{ij}$. As shown in **Fig. 4b**, under -3%⩽$\varepsilon$⩽3%, the $d_{11}$ values in ML MoS$_2$ can be linearly changed by ~50%. The extremely $\varepsilon$-sensitive $d_{11}$ prevents MoS$_2$ (and other 2D NLO materials) for stable NLO applications [5,19]. Surprisingly, under the same $\varepsilon$ ranges, the $d_{16}$ values in ML BSO can only be slowly changed by ~10%, while the $d_{23}$ values can even maintain constant. This unexpected robustness of SHG intensities originates from the unusual elastic property, *i.e.*, near-zero Poisson's ratios, in ML BSO. In response to an $\varepsilon$, the BO$_4$ and SO$_4$ tetrahedra prefer to rigidly rotate, with negligible tetrahedral distortions (Table S5 [28]). As shown in **Fig. 4c**, under a negative (positive) $\varepsilon$, the neighboring SO$_4$ and BO$_4$ tetrahedra rotate clockwise (anticlockwise) and anticlockwise (clockwise) along $c$ axis, respectively, which can simultaneously decrease (increase) the angle $\theta$ between them. It is found that $\theta$ has a linear dependence of $\varepsilon$ (Fig. S7 [28]), consequently resulting in the linear $\varepsilon$-dependence of $P_a$ ($a$-axis component of macroscopic $P$, originated from the polarity of tetrahedra). Meanwhile, as shown in **Fig. 4c**, the $P_b$ and $P_c$ are $\varepsilon$-independent. Therefore, the $d_{23}$ ($\chi^{(2)}_{bcc}$) values can be invariable under different $\varepsilon$, because of the $\varepsilon$-independent $P_b$ and $P_c$. On the other hand, the $d_{16}$ ($\chi^{(2)}_{aab}$) values slowly increase as $\varepsilon$ increases, due to the linearly increased $P_a$ and $\varepsilon$-independent $P_b$, preventing the rapid changes of $d_{16}$ values under strains. Interestingly, besides of $d_{ij}$, it is found that the $E_g$ of ML BSO are also robust (<7%) against large $\varepsilon$, in contrast to that of MoS$_2$ (~50%) (Fig. S8 [28]). Again, this is because the rigid tetrahedra can largely maintain the original bonding-antibonding separation strengths in BSO under strains (**Fig. 3b**). Therefore, the unusual robustness of SHG response (especially for $d_{23}$) and DUV $E_g$ make ML BSO suitable for various applications under extreme conditions.

In conclusion, we have successfully demonstrated that the layered BSO oxide can exhibit both strong SHG effects and ultrawide DUV bandgap, comparable to the best known DUV NLO crystals. Interestingly, the SHG intensities in BSO systems are tunable by the layer thickness but robust against large strain effects, differing from other known 2D NLO materials. As the first discovered 2D DUV NLO system, BSO could provide an unprecedented platform to explore the DUV NLO physics and device applications in ML thickness even under extreme conditions.

**Acknowledgements:** The authors thank Drs. S. H. Zhang and F. Liang for helpful discussion. This work is supported by the NSFC (Grant Nos. 11704023, 51890864, 11634003), NSAF (Grant No. U1930402) and the Science Challenge Project (Grant No. TZ2016003).

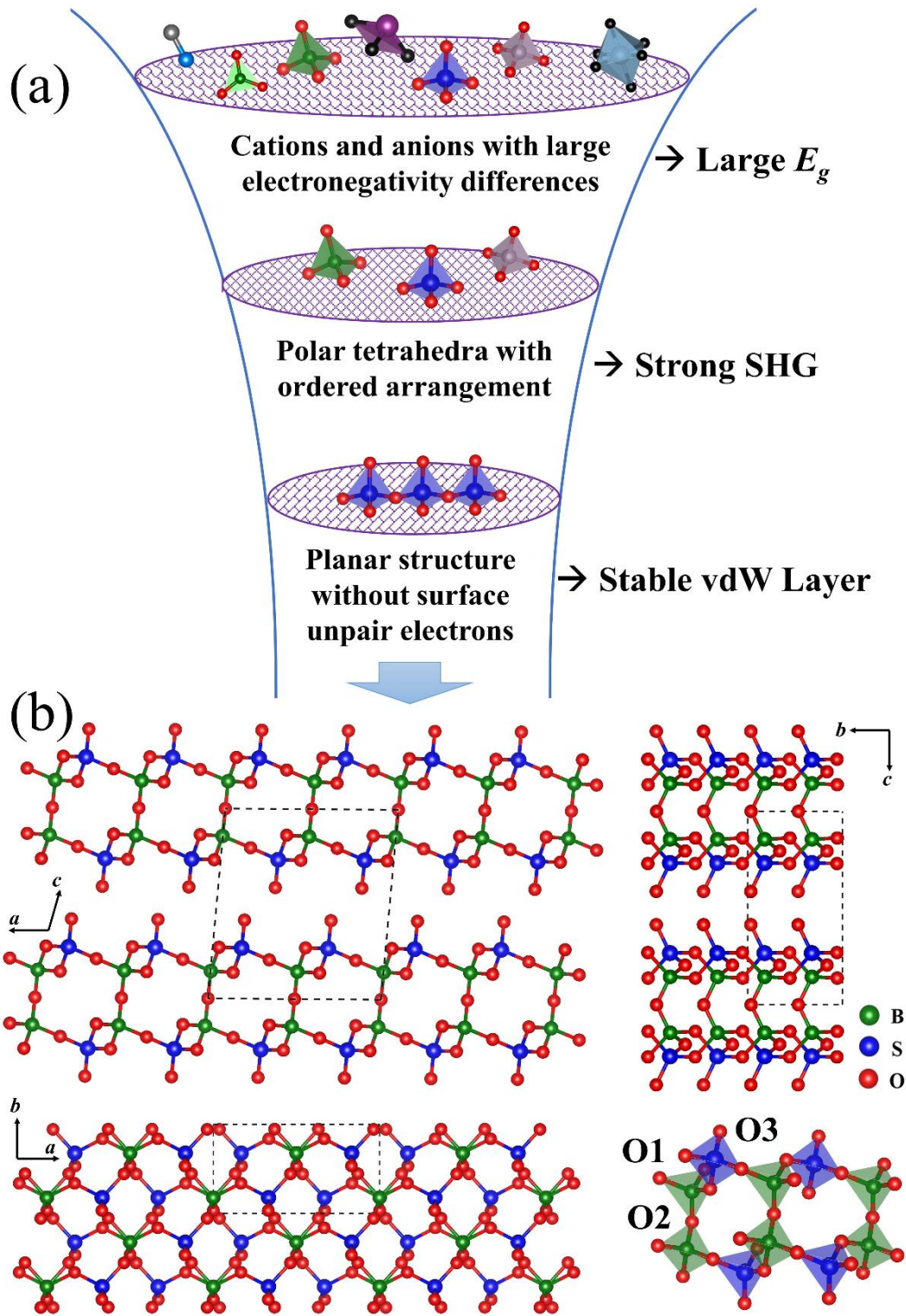

FIG. 1. (a) Procedures for screening DUV NLO vdW layered materials from ICSD, in terms of three criteria. (b) Side and top views of layered BSO bulk. Right-bottom: ordered arrangement of BO$_4$ and SO$_4$ tetrahedra forming BSO planar structure. Unitcells are marked by the black-dashed lines.

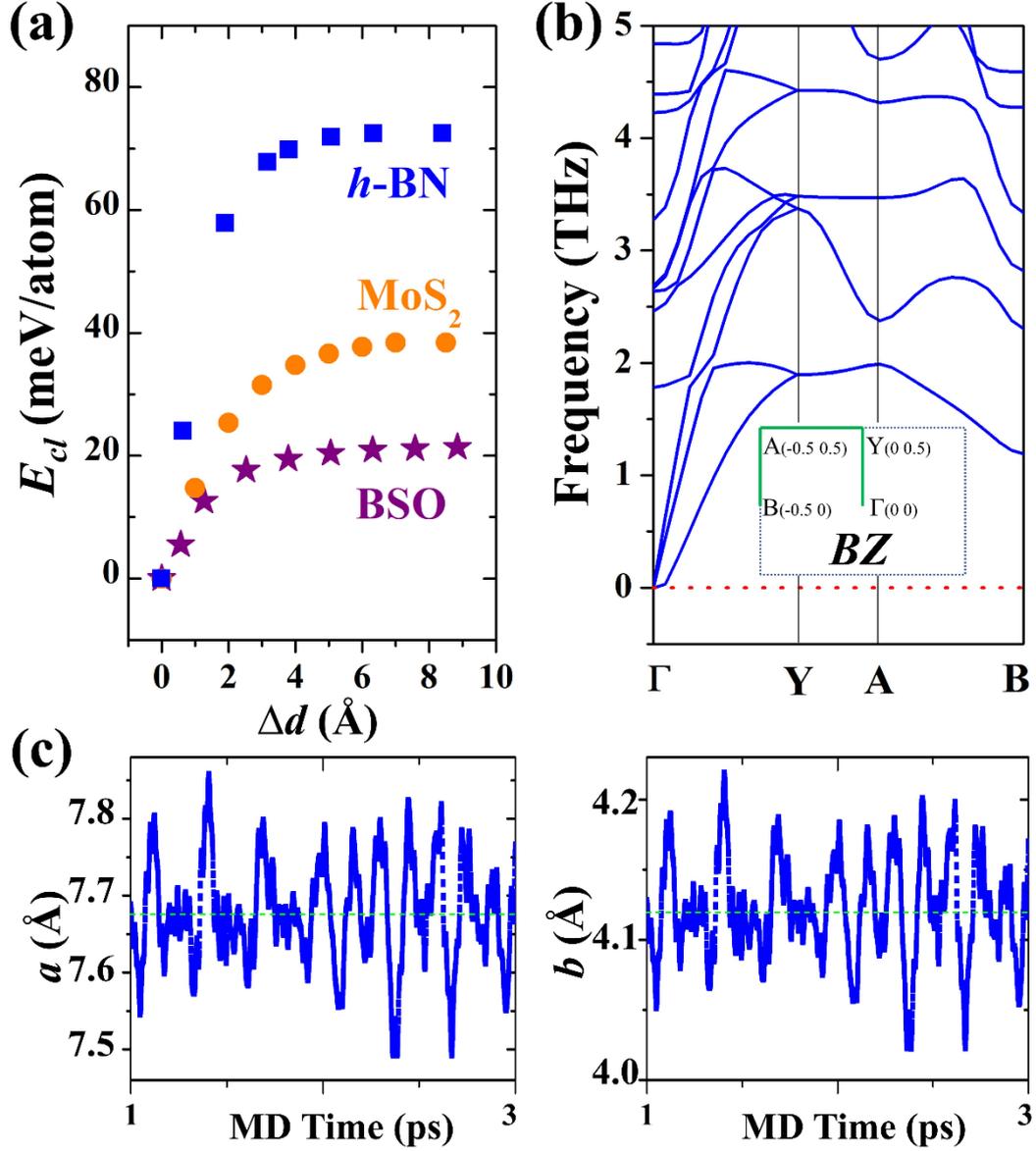

FIG. 2. (a) Formation energies of *h*-BN, MoS$_2$, and BSO as a function of interlayer distance changes $\Delta d$. (b) Phonon spectrum of ML BSO along high-symmetry *k*-paths in BZ. (c) Lattice constants *a* and *b* as a function of MD time from 1 to 3 ps. Green-dashed lines indicate the average lattice constants in MD simulations.

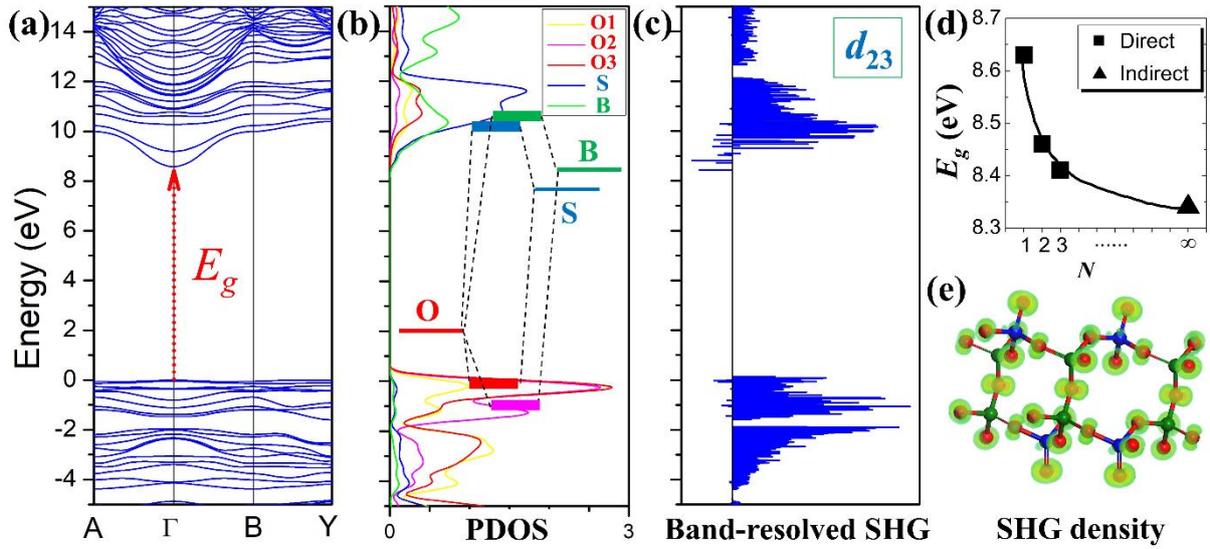

FIG. 3. (a) Band structure, (b) partial density of states (PDOS), and (c) band-resolved SHG ($d_{23}$) contributions of ML BSO, respectively. Fermi level is set to zero. Inset of (b): schematic plotting of cation-anion orbital-coupling forming the band edges and $E_g$ of BSO. (d) $E_g$ of BSO systems as a function of the thickness $N$. (e) SHG-weighted density of $d_{23}$ contributed by the atomic orbitals in ML BSO.

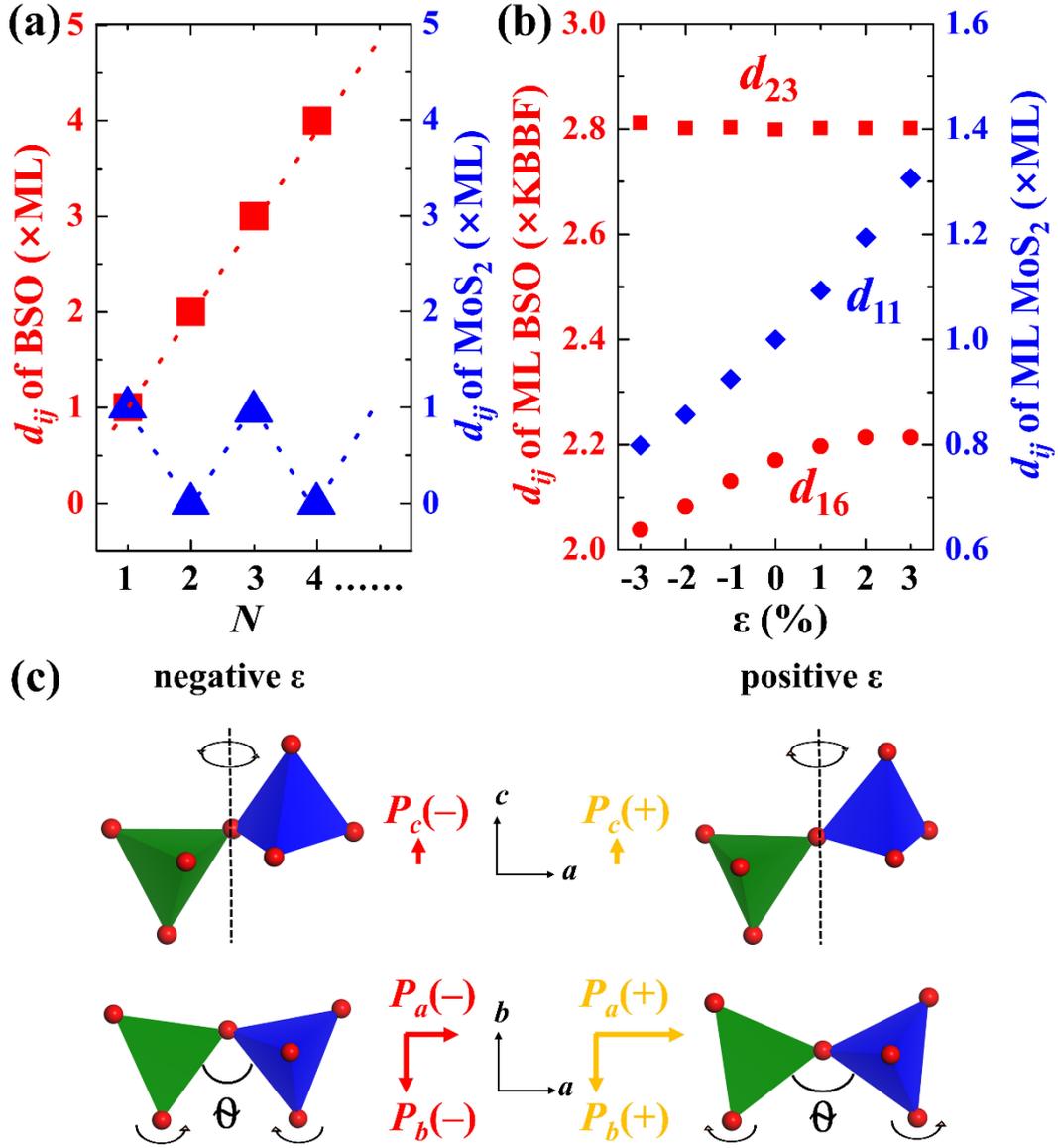

FIG. 4. (a) SHG coefficients $d_{ij}$ of BSO and MoS$_2$ as a function of $N$. (b) SHG coefficients $d_{ij}$ of ML BSO and MoS$_2$ as a function of biaxial strain ($\varepsilon$) from -3% to 3%. (c) Schematic plotting of the rigid rotations of neighboring BO$_4$ (green) and SO$_4$ (blue) tetrahedra under the negative and positive strains. Rotation axes are marked by dashed-lines. $a$-, $b$-, and $c$-axis components of macroscopic $P$, induced by the polarity of tetrahedra, are denoted as $P_a$, $P_b$, and $P_c$, respectively.